\documentclass[%
 aip,
 amsmath,amssymb,
 reprint,%
]{revtex4-1}

\usepackage{graphicx}
\usepackage{dcolumn}
\usepackage{bm}

\usepackage[utf8]{inputenc}
\usepackage[T1]{fontenc}
\usepackage{mathptmx}
\usepackage{etoolbox}

\makeatletter
\def\@email#1#2{%
 \endgroup
 \patchcmd{\titleblock@produce}
  {\frontmatter@RRAPformat}
  {\frontmatter@RRAPformat{\produce@RRAP{*#1\href{mailto:#2}{#2}}}\frontmatter@RRAPformat}
  {}{}
}%
\makeatother
\begin{document}

\preprint{AIP/123-QED}

\title[Plasma-generated nitrates as fertilizer for turf grass]{Plasma-generated nitrates as fertilizer for turf grass}
\author{Christina Sze}
\affiliation{Stanford University Department of Mechanical Engineering, Stanford, USA.}
\author{Benjamin Wang}
\affiliation{Stanford University Department of Mechanical Engineering, Stanford, USA.}
\author{Jiale Xu}
\affiliation{Stanford University Department of Electrical Engineering, Stanford, USA.}
\author{Juan Rivas-Davila}
\affiliation{Stanford University Department of Electrical Engineering, Stanford, USA.}
\author{M.A. Cappelli}
\affiliation{Stanford University Department of Mechanical Engineering, Stanford, USA.}

\date{\today}

\begin{abstract}

We investigated the use of plasma-fixated nitrogen, which produces nitrates ($NO_3^{-}$) in water, as a possible nitrogen fertilizer for recreational turf such as ryegrass and bentgrass. Experiments were carried out to study the effects of nitrate concentration on growth, the further effects of adding phosphorous (P) and potassium (K) to the plasma nitrated solution to make an N-P-K complete fertilizer, and to compare the efficacy of plasma-fixated nitrogen to sodium nitrate ($NaNO_3$) and potassium nitrate ($KNO_3$). The results indicate that the growth and biomass of the plants were strongly dependent on the concentration of the plasma-fixated nitrogen. Adding P-K to the plasma-fixated nitrogen improved grass growth. Grass that was supplied plasma-fixated nitrogen had improved growth compared to those supplied with equal amounts of $NaNO_3$ and $KNO_3$. This work highlights the potential use of plasma-fixated nitrogen as a fertilizer source for commonly used turf grass.

\end{abstract}

\maketitle

\section{Introduction}

Nitrogen is a key component of fertilizer and is essential to plant growth and health. Currently, most nitrogen in fertilizers is supplied as a nitrate, $NO_3^-$, bound to ammonia ($NH_4$) that is manufactured using the Haber-Bosch process\cite{Ingels2015a}. However, the Haber-Bosch process is energy-intensive and contributes to greenhouse gas emissions. 1\% - 2\% of global energy consumption and 1.44\% of carbon dioxide ($CO_2$) emissions are attributed to this ammonia production\cite{Kyriakou2020}. The harmful consequences of traditional nitrogen production illustrate the need for alternative processes for producing fertilizers. In this paper, we examine the efficacy of plasma-fixated nitrogen as a more sustainable fertilizer option, specifically for the fertilizing of commonly grown turf grass.

A gas discharge plasma, sometimes referred to as the "fourth state of matter", when generated in a molecular gas such as air, is composed of electrons, positive and negative ions, excited and neutral atoms, and other reactive species and radicals. Plasma fixation of nitrogen in water, sometimes referred to as "plasma-activated water," can be generated via air plasma treatment either above or directly in water (i.e., the water is exposed to a plasma stream, or a plasma is generated directly in the water). This results in the formation of numerous dissolved and chemically active species, often generalized as reactive oxygen and nitrogen species (RONS) \cite{Li2018a}. The dissolved reactive species can include nitrates, which are naturally present in soil and can be absorbed by plants as a source of nitrogen to promote plant growth. In this paper, we are specifically interested in the fixation of nitrate ions (plasma-fixated nitrogen), produced in water from exposure to air plasmas, for use as an exogenous fertilizer source for commonly used turf grass. 

The application of plasma-fixated nitrogen in agriculture has recently gained much attention because of its various interesting properties and potential for sustainable production. Plasma activated water has been shown to enhance seed germination \cite{Guo2021}, plant growth \cite{Thirumdas2018,Stoleru2020}, and antiseptic properties as a result of microbial reduction\cite{Zhao2020}. However, most applications to agriculture have focused on food production and crops such as corn\cite{Lamichhane2021}, barley\cite{Gierczik2020}, and fresh produce such as lettuce\cite{Stoleru2020}. In this paper, we examine the use of plasma-fixated nitrogen in the fertilization of turf grass. Turf grass is considered to be one of the largest irrigated crops in the United States, covering a greater surface area than even irrigated corn. Turf grass is commonly used in residential and commercial lawns, golf courses, and recreational and sports fields\cite{Selhorst2013}. As such, turf grass has the potential to sequester a large amount of excess carbon from the atmosphere and reduce greenhouse gas emissions\cite{Zirkle2011,Selhorst2013}. The use of sustainable energy in the production of plasma-fixated nitrogen for turf grass applications would further reduce the carbon footprint of turf grass.

\section{Materials and methods}

The focus of this study is to better understand the efficacy of plasma-fixated nitrogen as a fertilizer for ryegrass and bentgrass. Prior to use, seeds of these two grasses were kept at room temperature in cool, dry conditions. Coconut coir was used as a soil growth medium. 

Air plasmas were generated using an atmospheric dielectric barrier discharge (DBD) reactor\cite{Guo2021,Growth2020,Lamichhane2021} operating at 23 kHz frequency and with a sinusoidal peak to peak voltage of 7 kV. In a typical experiment, the discharge drew 450 watts of input power. Industrial water contained in a glass container is placed close to the surface of the DBD (< 1 cm) and treated until the the plasma-fixated nitrogen solution reached a desired pH and nitrate level. 

The plasma-fixated nitrogen stock solution was diluted with industrial water to achieve solutions with lower nitrate concentrations. These solutions had dilution factors of 2:1, 5:1, 10:1, 20:1, and 100:1, respectively, and were compared for their efficacy in growth studies. The nitrate concentrations and pH of each solution were measured prior to administering the solutions, and the results are summarized in Table I. The nitrate concentration of the diluted plasma-fixated nitrogen solutions was measured using a nitrate ion-selective electrode (Vernier GDX-NO3) as nitrates as nitrogen ($NO_3-N$). The pH of the solution was measured using a glass bodied pH sensor (Vernier GDX-GPH).

\begin{table}
\caption{\label{tab:table4}$NO_3-N$ concentration and pH of selected dilutions of plasma-fixated nitrogen solution.}
\begin{center} 
\begin{tabular}{|c|c|c|}
Treatment&$NO_3-N$\hspace{0.1cm}(ppm)&\mbox{pH}\\
\hline
Water & 2.54 & 6.79 \\
100:1 & 2.86 & 6.21 \\
20:1 & 7.22 & 4.21 \\
10:1 & 14.56 & 3.51 \\
5:1 & 31.61 & 3.06 \\
2:1 & 85.71 & 2.63 \\
Stock & 168 & 2.36 \\

\end{tabular}
\end{center}
\end{table}

\subsection{Experimental Setup}

\begin{figure}[htbp!]
\centering
\includegraphics[width=\linewidth]{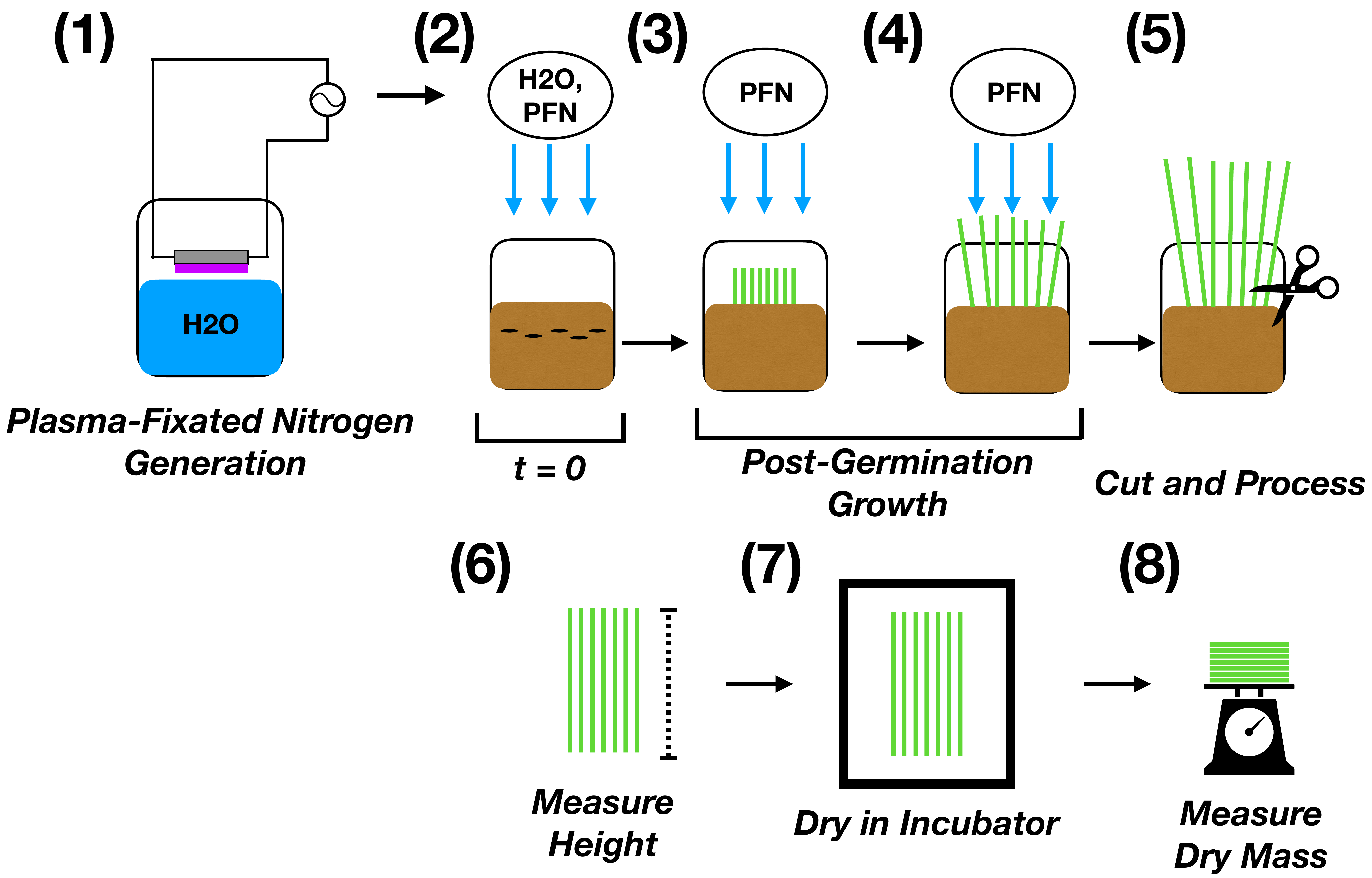}
\caption{The experimental setup for the grass growth studies. (1) Plasma-fixated nitrogen generated with DBD reactor. (2) Daily watering with water and plasma-fixated nitrogen during germination. (3) and (4) Daily watering with plasma-fixated nitrogen post-germination. (5) Grass cut and processed. (6) Height measurements recorded. (7) Grass dried in incubator for 2 days. (8) Dry mass measurements recorded.}
\label{fig:ExpSetup}
\end{figure}

The effect of (i) dilution on plasma-fixated nitrogens' growth effectiveness, (ii) the addition of phosphorous (P) and potassium (K) to undiluted plasma-fixated nitrogen solution, and (iii) plasma-fixated nitrogen performance against other sources of plant nitrogen are presented below. The general setup for all three studies is shown in Figure \ref{fig:ExpSetup}.

In the plasma-fixated nitrogen dilution study, 50 uniform ryegrass seeds were placed in 250 mL plastic containers on top of approximately 70 cubic centimeters of coco coir and covered with approximately 20 cubic centimeters of coco coir. Equal volumes of water and fertilizer were added to each respective container daily. The amount of nitrate added to each treatment group over time was recorded (Figure \ref{fig:exp1photonitrate}c), and final measurements were collected after 17 days following initial planting.

To test the effect of adding P-K to plasma-fixated nitrogen stock solution, liquid fertilizer containing only P-K (Liquid KoolBloom 0-10-10) was diluted by a factor of 1000 in industrial water. This solution was added to 168 ppm $NO_3-N$ of plasma-fixated nitrogen stock solution to form an N-P-K complete fertilizer solution. The diluted P-K fertilizer was also tested without the added plasma-fixated nitrogen in this study as a control. 25 uniformly spaced ryegrass seeds were placed in 250 mL plastic containers on top of approximately 250 cubic centimeters of coco coir and covered with approximately a quarter inch of coco coir, or about 40 cubic centimeters of coco coir. Equal volumes of water and fertilizer solution were added daily to each respective container. The amount of nitrate added to each treatment group over time was recorded (Figure \ref{fig:exp2photonitrate}b and d). Four replicates for each treatment group were used in this study. Measurements were recorded after 19 days following initial planting.

To compare the performance of plasma-fixated nitrogen against other sources of plant nitrogen, 5 seeds were placed in approximately 7 grams of coco coir with a quarter inch of coco coir over the seeds, and either 2.5 mL of water (control), 100 ppm $NO_3-N$ plasma-fixated nitrogen solution, 100 ppm sodium nitrate as nitrogen ($NaNO_3$, ASI Sensors), and 100 ppm potassium nitrate as nitrogen ($KNO_3$, LabChem) were added daily to each respective growth container until germination. After germination, equal volumes of the solutions were added daily to each respective treatment group. The amount of nitrate added to each treatment group over time was recorded (Figure \ref{fig:exp3photonitrate}). Six replicates for each treatment group were used. Final measurements were recorded after 13 days following initial planting.

For all of the growth studies, the seeds were placed in a custom climate-controlled (T = 22$^{o}$ C) plant growth chamber with 24-hour light exposure during the germination period and 14-hour light exposure during post-germination and plant growth.

\subsection{Growth analysis}

Plant growth was evaluated based on average grass height and final dry mass of each treatment group. Each blade of grass was cut at the crown of the plant and individually measured with its natural curvature using a caliper (Mitutoyo Absolute Digimatic 200mm). Grass samples were then dried in an incubator (VEVOR 20L RT-65) for 2 days at 40$^{o}$C before being weighed using a scale (Ohaus H-7293 Scout) to obtain dry mass measurements. Average dry mass measurements per plant were calculated by taking the dry mass and dividing by the number of germinated plants per replicate group.

Germination yield was calculated by dividing the number of germinated seeds by the total number of seeds in each cup using the following equation:

\[G = n_G / n_T \times 100\% \]

Here, $G$ is the germination percent yield, $n_G$ is the number of germinated seeds in each cup, and $n_T$ is the total number of seeds in each cup.  

\subsection{Statistical analysis}

Two-tailed, two-sample heteroscedastic t-tests were run to determine whether there was a statistically significant difference in the average heights of the treatment groups (significance level $\alpha < $ 0.05).

\section{Results and Discussion}

\subsection{Increased plasma nitrate concentration enhances grass growth}

\begin{figure*}[hbt!]
\centering
\includegraphics[width=\linewidth]{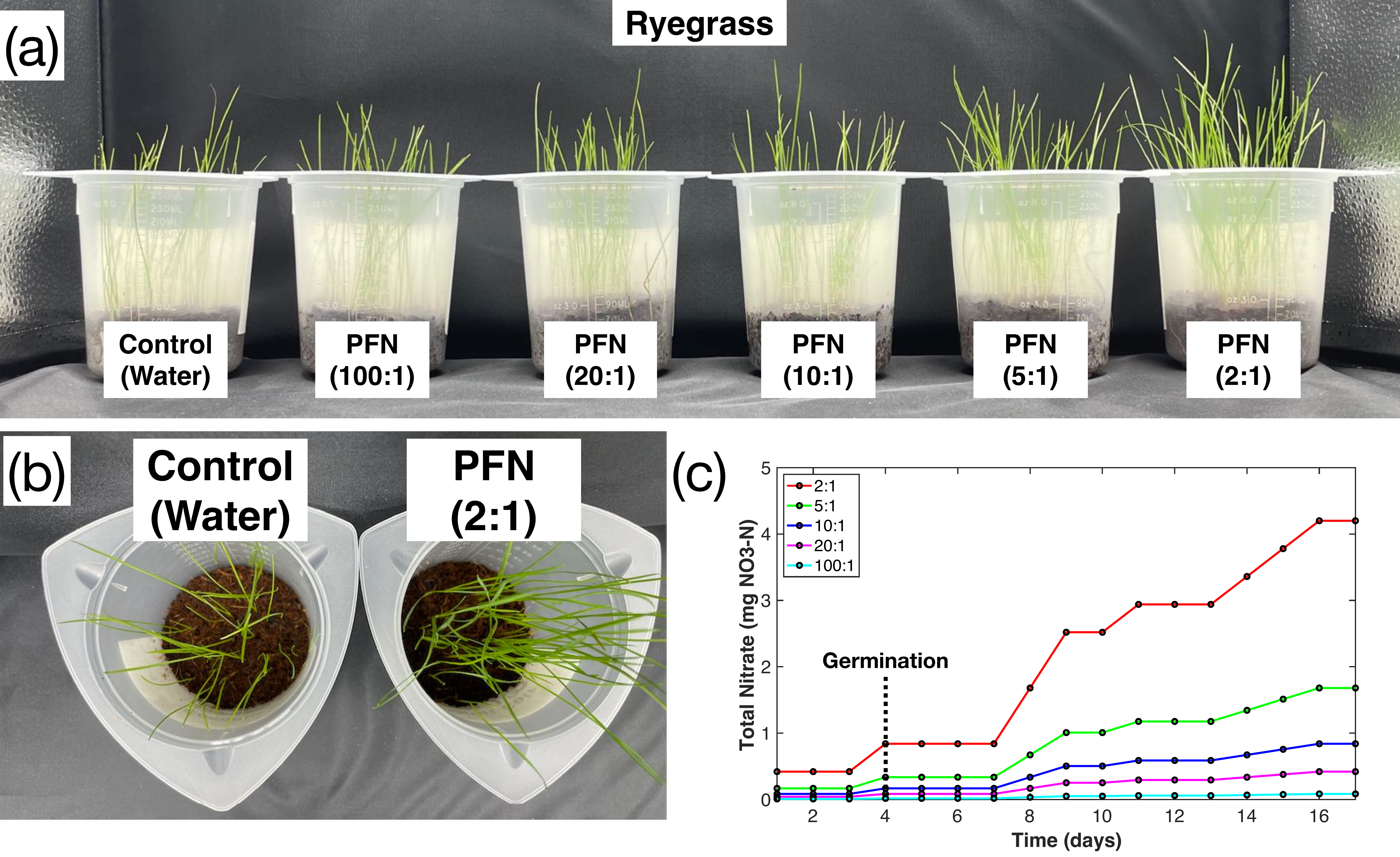}
\caption{(a) Photograph of ryegrass growth results after 17 days for various dilutions of plasma-fixated nitrogen. (b) Top view image comparing control (left) and 2:1 dilution of plasma-fixated nitrogen fertilizer (right)'s effect on the growth of ryegrass. (c) Total nitrate schedule for each fertilizer dilution.}
\label{fig:exp1photonitrate}
\end{figure*} 

The results of the plasma-fixated nitrogen's effect on turf grass growth are shown in Figure \ref{fig:exp1photonitrate}. After 17 days of growth, the turf grass with higher concentrations of plasma-fixated nitrogen (20:1, 10:1, 5:1, 2:1 dilution ratio from the 168 ppm $NO_3-N$ stock) were visibly taller, thicker, denser, and greener compared to the groups that received only water (control) or dilute amounts of the fertilizer (100:1 dilution). Growth improved with regard to height, color, and thickness as fertilizer concentration increased (Figure \ref{fig:exp1photonitrate}a). This difference is most clearly illustrated when comparing the control (water) and the 2:1 treatment group in Figure \ref{fig:exp1photonitrate}b, which had the greatest difference in total nitrate added over the growth period.

\begin{figure*}[htb!]
\centering
\includegraphics[width=\linewidth]{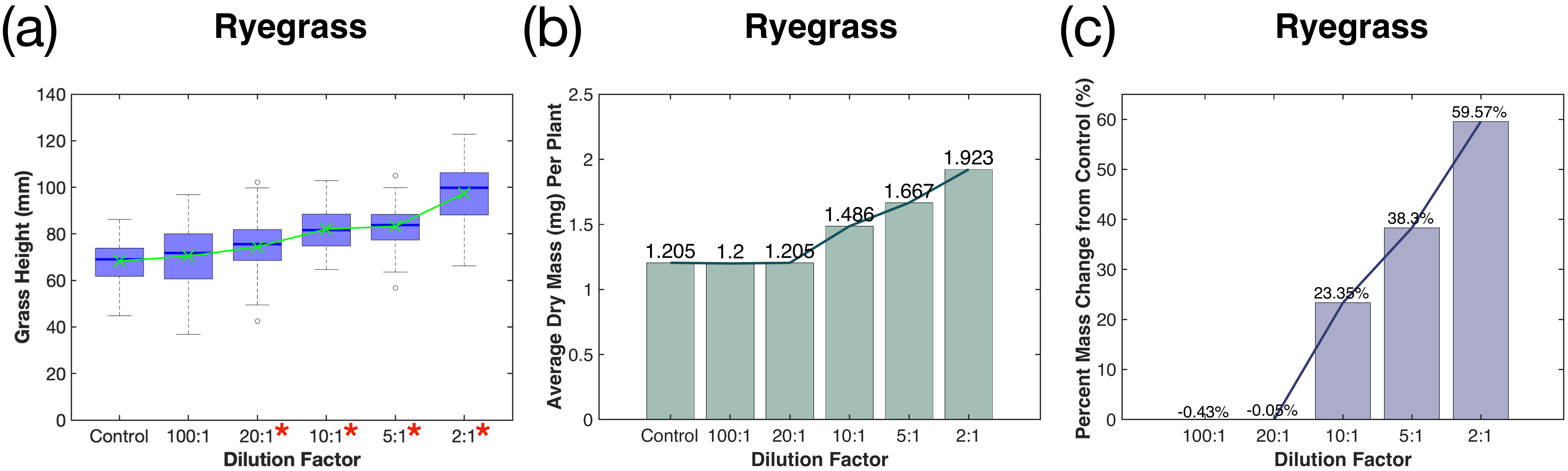}
\caption{(a) Ryegrass height as a function of plasma-fixated nitrogen levels. '*' indicates a statistically significant different distribution from the control. The light green 'x' indicates the mean grass height for each dilution group. (b) Average ryegrass dry mass per plant as a function of nitrate levels. (c) Percent change in dry mass per plant as a function of added nitrates amounts.}
\label{fig:exp1heightmass}
\end{figure*}

The mean grass height of every treatment group that received diluted plasma-fixated nitrogen fertilizer, except for the 100:1 dilution, was found to be significantly greater ($\rho < 0.05$) from the mean height value measured for the control group. The height distributions of the treatment groups are shown in a box plot in Figure \ref{fig:exp1heightmass}a with the mean height plotted in green. As the fertilizer and $NO_3-N$ concentration increases, the mean height and the height distribution also increases, particularly for the highest concentration (2:1 dilution) treatment group. Increasing fertilizer concentration was also shown to increase grass biomass (Figure \ref{fig:exp1heightmass}b). The greatest effect was seen in the 5:1 and 2:1 dilution treatment groups, which had a 38.3\% and 59.57\% increase in dry mass when compared to the control, respectively (Figure \ref{fig:exp1heightmass}c). It is apparent that adding even small amounts of plasma-fixated nitrogen (4.20 mg and 1.68 mg for the 2:1 and 5:1 groups) dramatically increased biomass. It is noteworthy that the germination yield was not affected by the application of the plasma-fixated nitrogen. The value of $G$ for the control group was 78\%, and $G$ values for the diluted fertilizer groups were similar (G = 80\%, 88\%, 74\%, 78\%, and 78\% for the 100:1, 20:1, 10:1, 5:1, and 2:1 dilutions, respectively).

Recent studies have found that the addition of nitrogen to Aceraceae, a deciduous shrub, improved plant growth and health through improved root morphology\cite{Razaq2017}. While we have not examined the physical morphology of our grass roots in detail, we conjecture that the benefits of added nitrogen, seen here as an increase in height and biomass, may also be the result of changes to root morphology. It is noteworthy that the more concentrated, acidic fertilizers (2:1 and 5:1) did not appear to "burn" the grass or its root system, suggesting the coco coir acted as an effective buffer to counteract the acidity of the fertilizer, in addition to the relatively low amount of solution that was added compared to the growth medium.

In studying the effect of adding plasma-fixated nitrogen on the growth of the plants, we considered two other quantitative measures for comparison that represent an efficiency factor for the nitrates on growth. The first is the ratio of the total added nitrates (as nitrogen) for the $i^{th}$ sample, $(NO_{3}-N)_{i}$, measured as mg of nitrogen, to the mean grass height of the $i^{th}$ sample beyond that of the control, i.e.,

$$\Gamma_{i} = \frac{(NO_{3}-N)_{i}}{grass\;height_{i} - grass\;height\;control}$$

The second is the ratio of the total added nitrates as nitrogen to the dry mass of the $i^{th}$ sample beyond that of the control, i.e.,

$$\xi_{i} = \frac{(NO_{3}-N)_{i}}{grass\;dry\;weight_{i} - grass\;dry\;weight\;control}$$

\begin{figure*}[htb!]
\centering
\includegraphics[width=\linewidth]{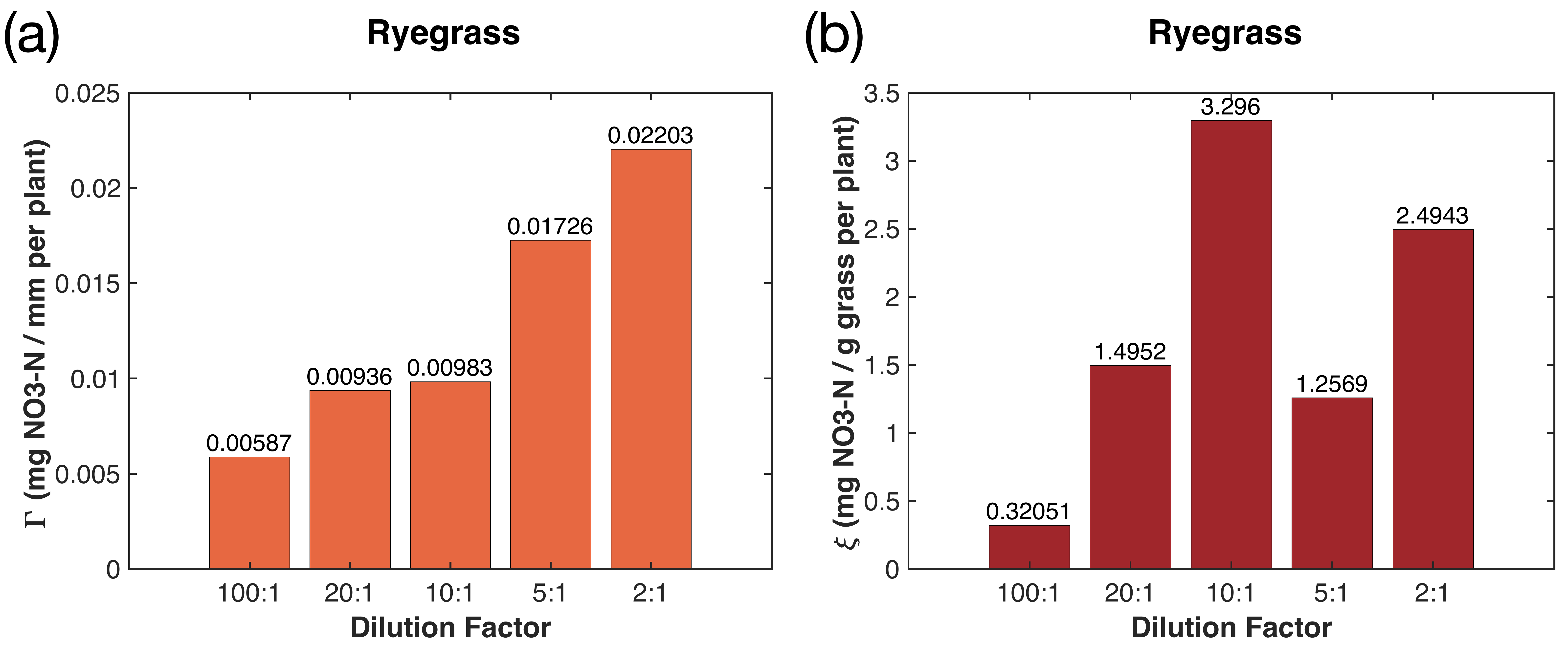}
\caption{(a) Effect of dilution factor on $\Gamma$, the total amount of mg of nitrates added during the experiment per mm of plant height. (b) Effect of dilution factor on $\xi$, on total amount of mg of nitrates added during the growth per gram of dry mass of plant.}
\label{fig:exp1diffheightmass}
\end{figure*}

These two ratios allowed us to further quantify the efficacy of the plasma nitrated water. Figure \ref{fig:exp1diffheightmass}a shows that the 100:1 dilution case only required 0.00587 mg of $NO_3-N$ per additional millimeter of height, indicating that the 100:1 dilution had the most efficient usage of nitrogen in increasing height. This efficiency of nitrogen usage decreased as $NO_3-N$ concentration increased, suggesting that more nitrogen was needed to increase grass height by the same amount. This inverse relationship between nitrate concentration and nitrate efficiency in increasing height suggests that it may be more economical to fertilize more often with smaller amounts of nitrogen, if grass height is desirable. Figure \ref{fig:exp1diffheightmass}b shows the amount of nitrates required to increase grass biomass by 1 mg. Similar to \ref{fig:exp1diffheightmass}a, the 100:1 dilution had the most efficient usage of nitrogen in increasing biomass since it only required 0.32051 mg of $NO_3-N$ per additional mg of biomass (on a per plant basis). However, the inefficiency of nitrogen usage in increasing biomass appears to peak at the 10:1 dilution and again at the 2:1 dilution of the fertilizer. 

Determining the optimal dilution for increasing nitrogen fertilizer efficiency is important toward reducing fertilizer costs and runoff, and adjusting the amount of nitrates we add to the plants over time is crucial for optimizing growth\cite{Growth2020}. While the 100:1 dilution had the best efficiency for increasing height and mass, the growth rate was slow, and there was not a significant difference (significance level $\alpha < 0.05$) in height from the control group over the 17-day growth period. Therefore, the results suggest that the 20:1 dilution may be the best in terms of maximizing growth rate and nitrogen efficiency regarding height. Similarly, the 5:1 dilution is likely the best in terms of maximizing growth rate and nitrogen efficiency regarding biomass. Further exploration is needed to determine the precise relationship between nitrogen efficiency in increasing mass and nitrate concentration, and particularly why certain dilutions like the 10:1 dilution were very inefficient in using the added nitrogen to increase mass.

\subsection{Addition of phosphorus and potassium to plasma-fixated nitrogen}

\begin{figure*}[!htbp]
\centering
\includegraphics[width=\linewidth]{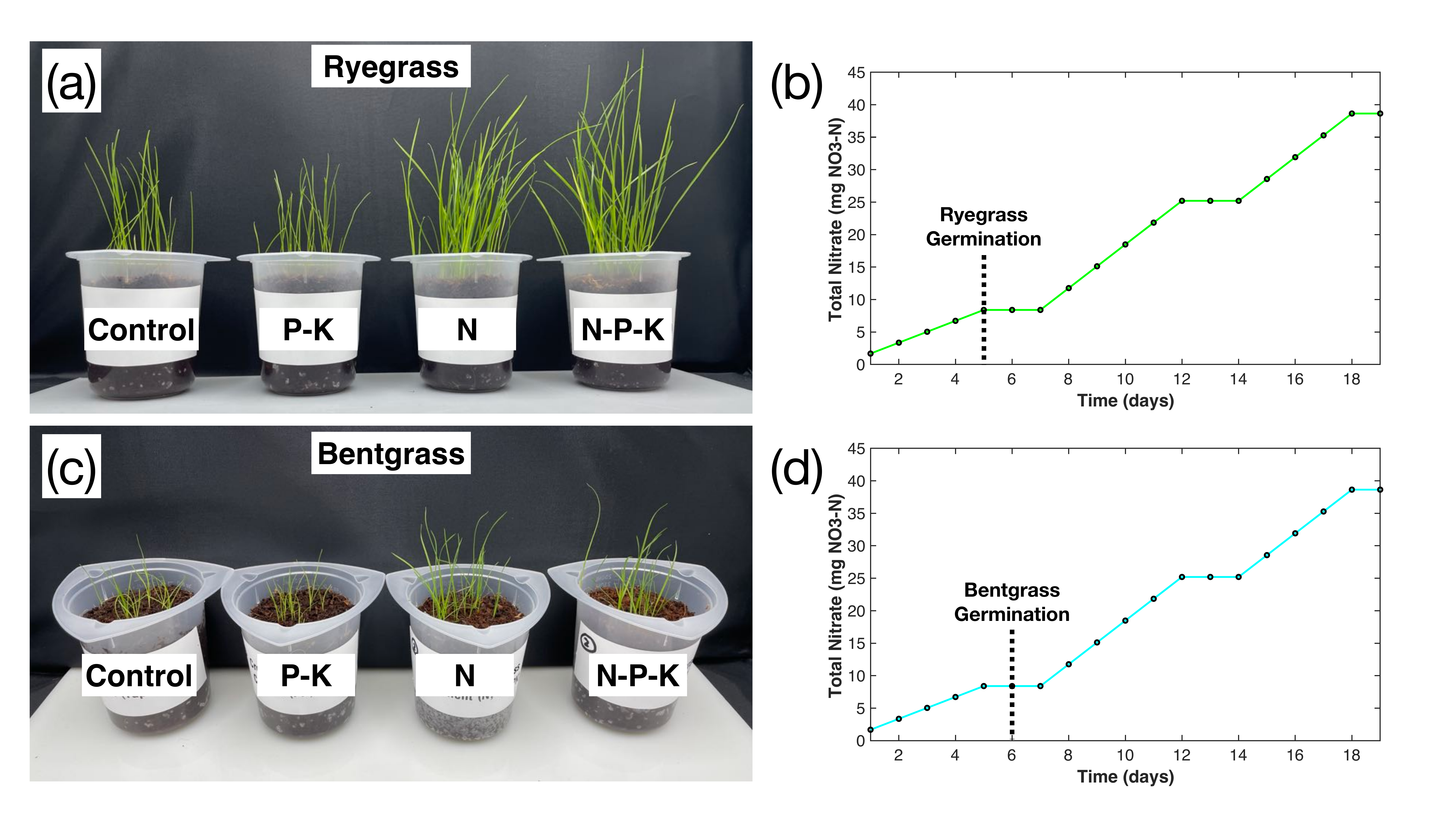}
\caption{The effect of watering ryegrass (a) and bentgrass (c) with tap water, P-K solution, N solution, and N-P-K solution. Photograph of sample plants watered with various solutions and the control treatment (industrial water) depicting growth on day 19. (b, d) Total nitrates given to N and N-P-K groups during growth cycle. Abbreviation P-K (Phosphorus Potassium), N (Plasma-Fixated), N-P-K (Nitrogen (Plasma-Fixated) Phosphorus Potassium).}
\label{fig:exp2photonitrate}
\end{figure*}

Phosphorus (P) and potassium (K) play a vital role in the growth of plants, and make up the two other essential macronutrients for plant growth \cite{Razaq2017,Xu2020}. This study investigated the addition of P-K to the plasma-fixated nitrogen solution in order to form a N-P-K complete fertilizer solution and better understand the effects of P-K on plasma-fixated nitrogen. The results of adding P-K to the industrial water control and the plasma-fixated nitrogen solution are shown in Figure \ref{fig:exp2photonitrate}. For both the ryegrass and the bentgrass, the P-K solution resulted in slightly shorter and thinner grass blades, while adding N-P-K plasma-fixated nitrogen solution visibly enhanced grass growth and moderately increased height (Figure \ref{fig:exp2photonitrate}a and c). Visually, the ryegrass and bentgrass treated with the N-P-K solution had thicker leaf blades and were more green in color compared to the ryegrass and bentgrass treated with the P-K solution and water (control). 

\begin{figure*}[htbp]
\centering
\includegraphics[width=\linewidth]{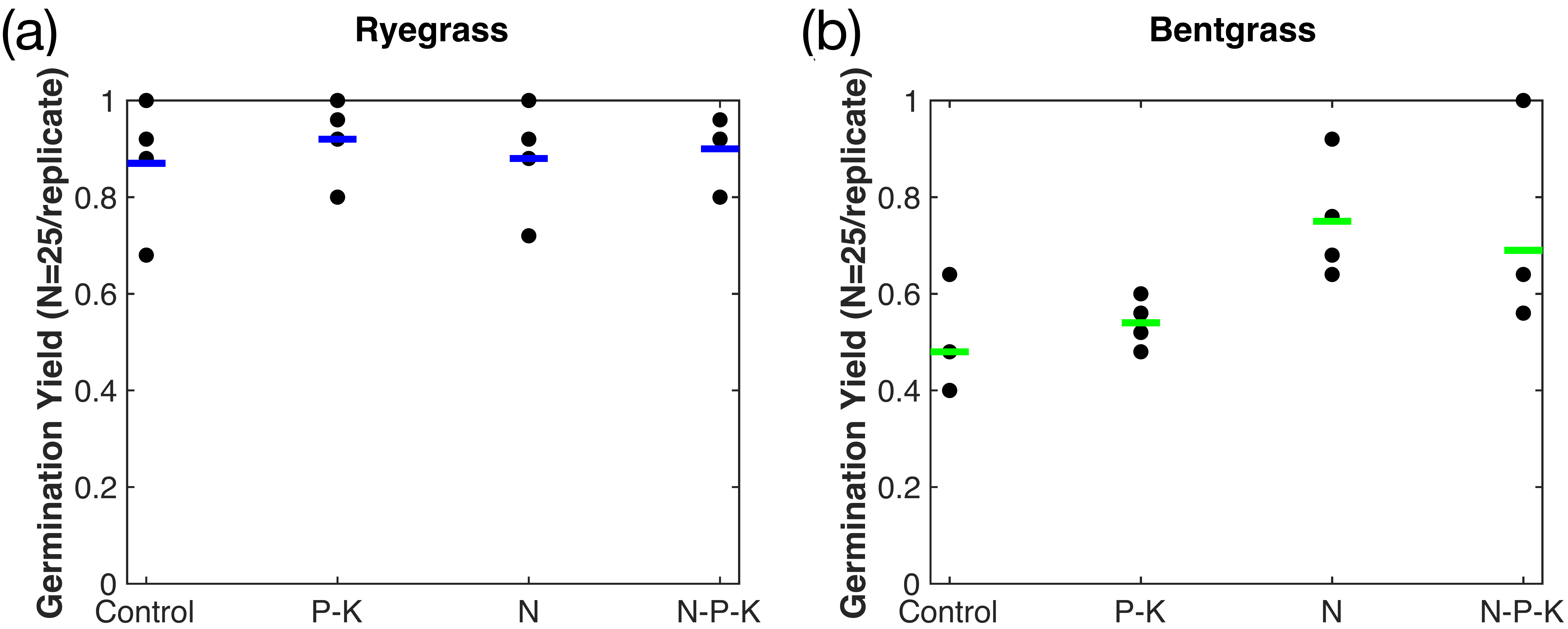}
\caption{Germination yield for (a) ryegrass and (b) bentgrass growth experiments, with N = 25 seeds per replicate groups (total of 4 replicate groups). The average germination yield is marked in blue (ryegrass) and green (bentgrass). Abbreviation P-K (Phosphorus Potassium), N (plasma-fixated), N-P-K (Nitrogen (plasma-fixated) Phosphorus Potassium).}
\label{fig:exp2germ}
\end{figure*}

Figure \ref{fig:exp2germ}a and Figure \ref{fig:exp2germ}b show the measured germination yield for both the ryegrass and bentgrass experiments, respectively. Ryegrass had an average germination yield of 87\% for the control, 92\% for the P-K solution, 88\% for the N solution, and 90\% for the N-P-K solution, as shown by the blue bars in Figure \ref{fig:exp2germ}a. Therefore, the addition of P-K did not noticeably affect germination yield between experimental groups for the ryegrass. The bentgrass germination yield had an average germination yield of 48\% for the control, 54\% for the P-K solution, 75\% for the N solution, and 69\% for the N-P-K solution, as shown by the green bars in Figure \ref{fig:exp2germ}b. The N and N-P-K treatment groups had higher germination yields compared to the control and P-K groups, suggesting that the plasma-fixated nitrogen improved bentgrass seed germination.

\begin{figure*}[!htbp]
\centering
\includegraphics[width=\linewidth]{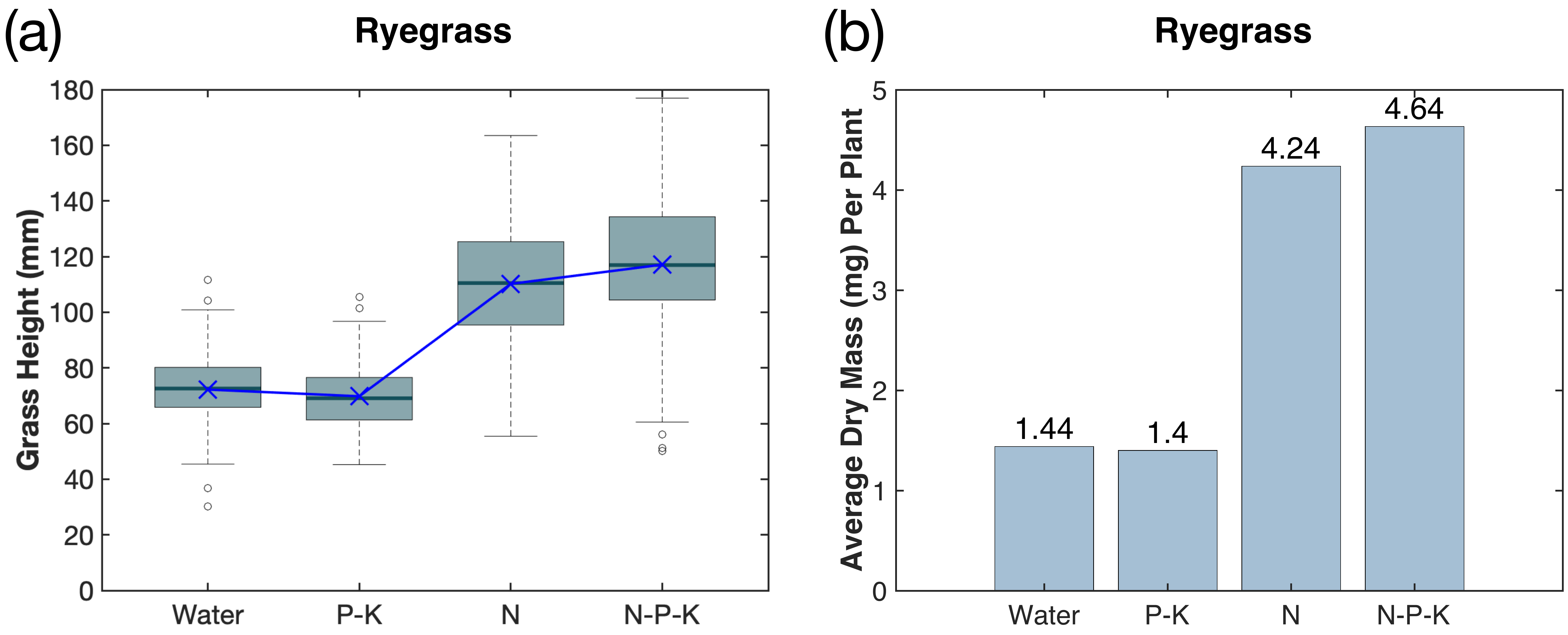}
\caption{Ryegrass (a) height and (b) average dry mass per plant with N = 25 seeds per replicate groups (total 4 replicate groups). The dark blue 'x' indicates the mean grass height for each fertilizer group. Abbreviation P-K (Phosphorus Potassium), N (Plasma-Fixated), N-P-K (Nitrogen (Plasma-Fixated) Phosphorus Potassium).}
\label{fig:exp2rgheightmass}
\end{figure*}

The addition of phosphorus and potassium to water and the plasma-fixated nitrogen solution had a small effect on ryegrass height, as shown in Figure \ref{fig:exp2rgheightmass}a. The mean height of each treatment group is plotted in dark blue. The P-K treatment group had a slightly lower height distribution compared to the control (water). However, this did not constitute a significant difference (significance level $\alpha < 0.05$). In contrast, the addition of P-K to the plasma-nitrated solution (N-P-K group) significantly increased the height distribution in comparison to the treatment group that received only the plasma nitrate solution (N group; $\rho < 0.05$). The N-P-K height distribution also has a wider range and a higher maximum height than the N height distribution, as well as a more symmetrical distribution.

Adding phosphorus and potassium to water alone had a small effect on ryegrass biomass. As shown in Figure \ref{fig:exp2rgheightmass}b, the control and P-K treatment group had approximately equal biomass with an average dry mass per plant of 1.44 mg and 1.40 mg, respectively. Similarly, the average dry masses per plant of the N and N-P-K treatment groups were 4.24 mg and 4.64 mg, respectively, suggesting the addition of P-K to the plasma-fixated nitrogen solution had a significant effect on increasing grass biomass.

\begin{figure*}[!htbp]
\centering
\includegraphics[width=\linewidth]{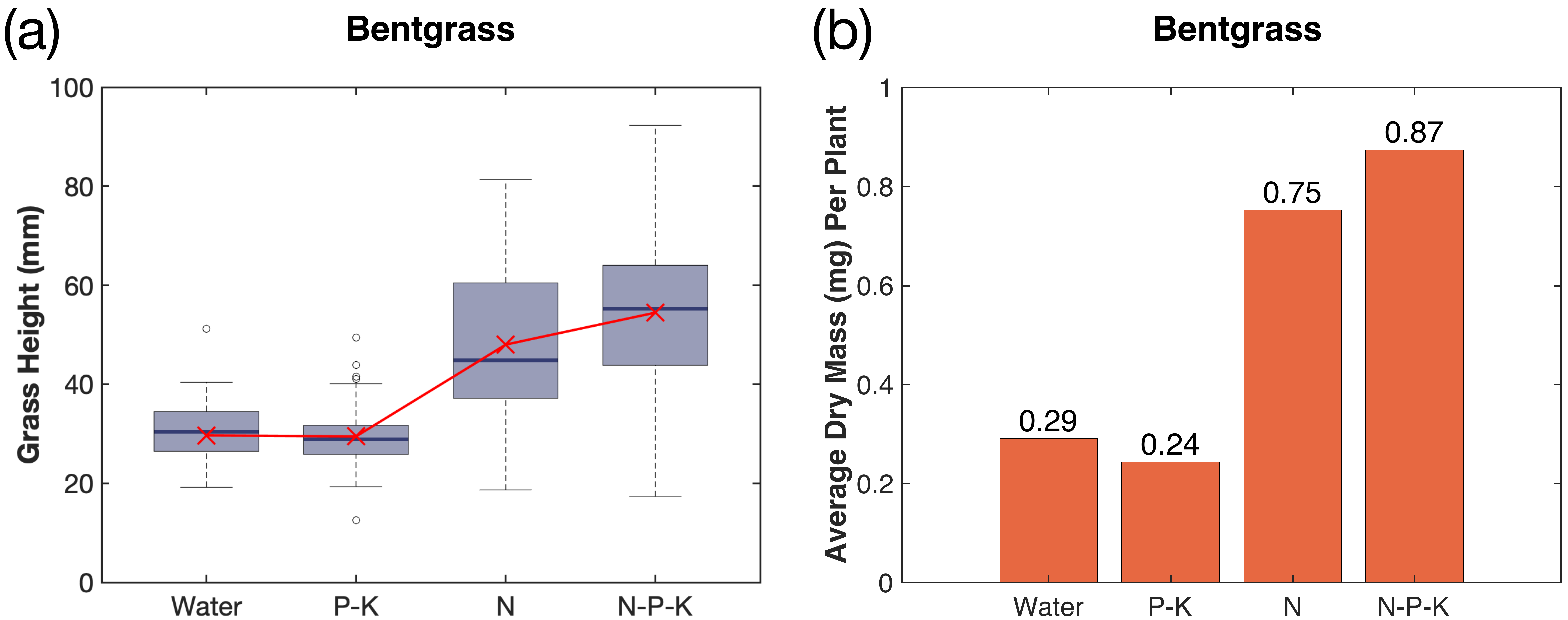}
\caption{Bentgrass (a) height and (b) dry mass with N = 25 seeds per replicate groups (total 4 replicate groups). The red 'x' indicates the mean grass height for each fertilizer group. Abbreviation P-K (Phosphorus Potassium), N (Plasma-Fixated), N-P-K (Nitrogen (Plasma-Fixated) Phosphorus Potassium).}
\label{fig:exp2bgheightmass}
\end{figure*}

The results from the creeping bentgrass seeds followed the same trend with regards to P-K's effect on height and biomass when added to water and plasma-fixated nitrogen. The control (water) and P-K groups had approximately the same height distribution, and the N-P-K treatment group had an increase in its median and range of height in comparison to the N treatment group (Figure \ref{fig:exp2bgheightmass}a). The mean height of each treatment group is plotted in red (Figure \ref{fig:exp2bgheightmass}a). The control and P-K groups had an average dry mass per plant of 0.29 mg and 0.24 mg, and the N and N-P-K treatment groups had an average dry mass per plant of 0.75 mg and 0.87 mg, indicating P-K had only a small effect on grass biomass (Figure \ref{fig:exp2bgheightmass}b).

Our study suggests that adding phosphorous and potassium without plasma-fixated nitrogen has a small but insignificant effect on decreasing plant growth, while adding phosphorous and potassium along with plasma-fixated nitrogen had a significant effect in enhancing plant height and biomass. These results are consistent with previous literature that showed adding phosphorous and potassium alongside nitrogen enhances plant growth\cite{Xu2020,Razaq2017}. 

\subsection{Comparison of plasma-fixated nitrogen to other nitrogen sources}

\begin{figure}[htbp!]
\centering
\includegraphics[width=\linewidth]{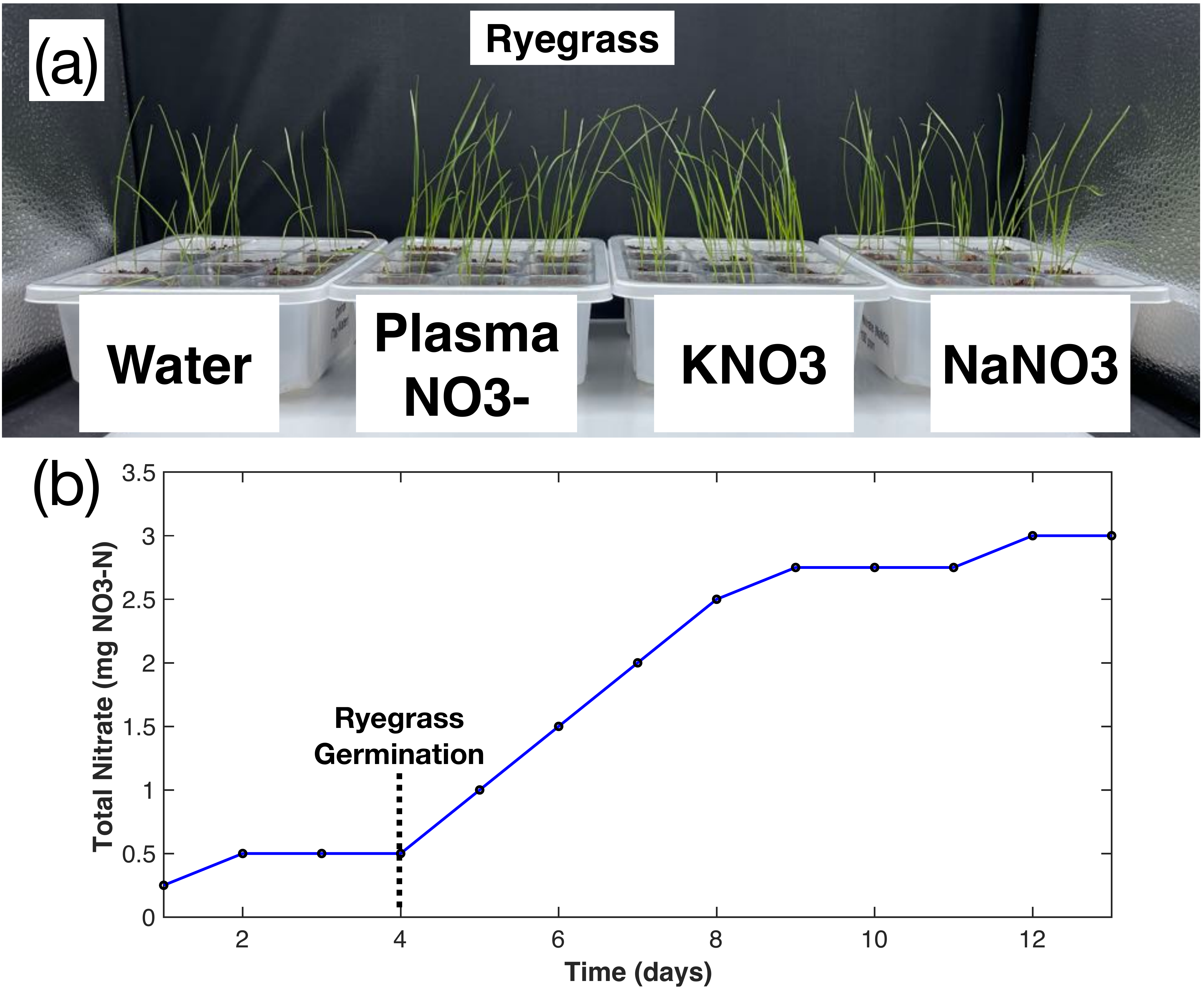}
\caption{(a) Photograph of ryegrass growth results after 13 days for comparison of control (water), plasma-fixated nitrogen solution (100 ppm $NO_3-N$), $KNO_3$, and $NaNO_3$ solutions (100 ppm $NO_3-N$). (b) Total nitrate schedule for each fertilizer.}
\label{fig:exp3photonitrate}
\end{figure}

\begin{figure*}[htbp!]
\centering
\includegraphics[width=\linewidth]{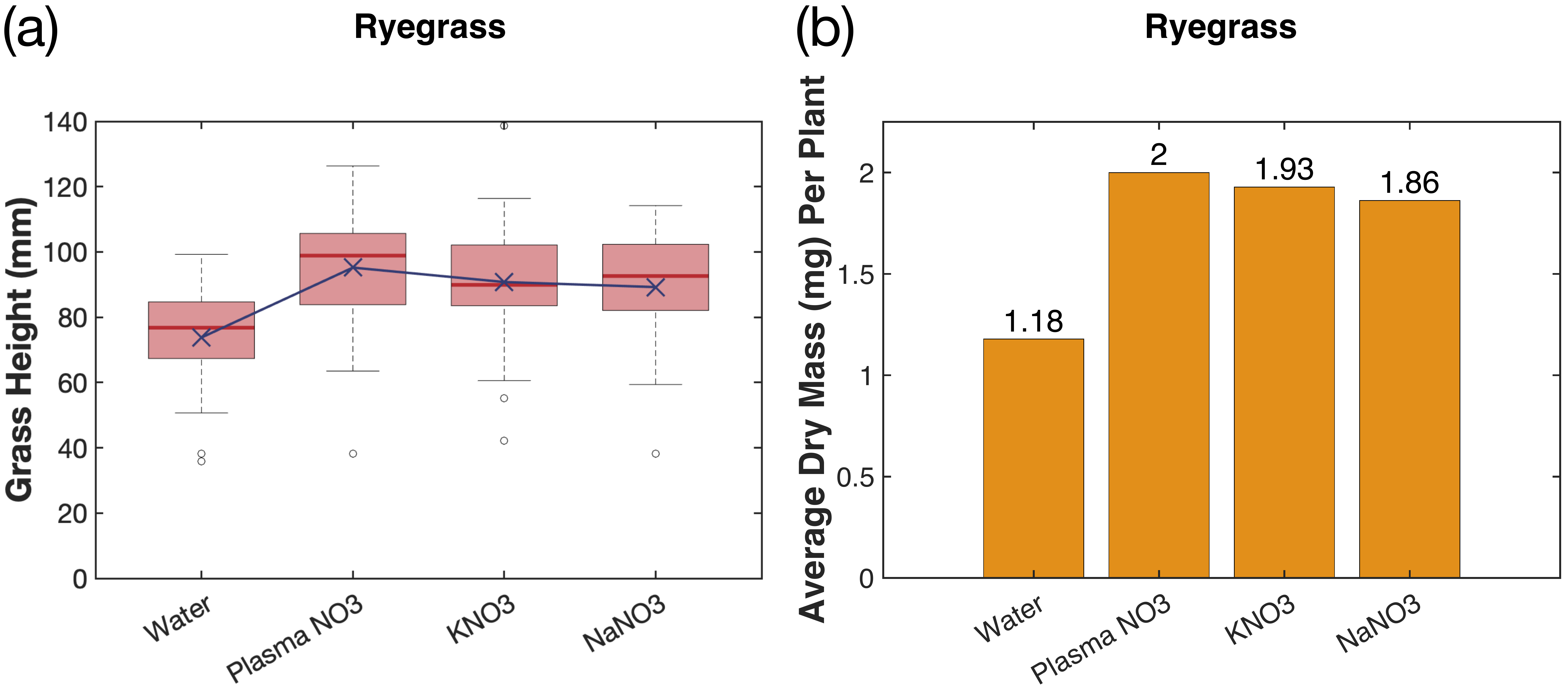}
\caption{Ryegrass (a) height and (b) average dry mass per plant with N = 5 seeds per replicate groups (total 6 replicate groups) for control (water), plasma-fixated nitrogen ($NO_3-N$), $KNO_3$, and $NaNO_3$ solutions. }
\label{fig:exp3heightmass}
\end{figure*}

Ryegrass grown using 100 ppm $NO_3-N$ plasma-fixated nitrogen had improved growth compared to ryegrass grown with $KNO_3$ and $NaNO_3$ at $NO_3-N$ concentrations of 100 ppm $NO_3-N$. The plasma-fixated nitrogen, $KNO_3$, and $NaNO_3$ groups were visibly the same in height, color, and blade thickness, and all showed significant improvement in growth compared to the control (see Figure \ref{fig:exp3photonitrate}a). These observations are reflected in the height distributions shown in Figure \ref{fig:exp3heightmass}a. The mean height of each treatment group is also plotted in purple in Figure \ref{fig:exp3heightmass}a. The plasma $NO_3^-$ had a higher mean height of 95.2 mm compared to $KNO_3$ (90.8 mm) and $NaNO_3$ (89.2 mm), although the level of significance in these differences is small ($\rho > 0.05$). The height distribution of the plasma-fixated nitrogen is wider than that of the $KNO_3$ and $NaNO_3$, and the plasma-fixated nitrogen had the highest maximum height.

The plasma-fixated nitrogen also produced a higher biomass compared to the $KNO_3$ and $NaNO_3$ solutions. The plasma $NO_3$ had an average dry mass per plant of 1.18 mg, while the $KNO_3$ and $NaNO_3$ groups had an average dry mass per plant of 1.93 mg and 1.86 mg, respectively (see Figure \ref{fig:exp3heightmass}b). Germination yield was also roughly equal across the different nitrogen sources ($G$ = 97\%, 93\%, and 96\% for the plasma $NO_3^-$, $KNO_3$, and $NaNO_3$, respectively). This shows that the nitrate source from the plasma-fixated solution has a higher performance in terms of growth and biomass compared to $KNO_3$ and $NaNO_3$ solutions of equal nitrate levels. The $KNO_3$ solution produced more biomass due to the availability of potassium nutrient, compared to the $NaNO_3$ solution. 

\section{Conclusions}

Three experiments were conducted in order to understand how effective plasma-fixated nitrogen fertilizer is in enhancing perennial ryegrass and creeping bentgrass growth. Dilutions of a stock solution of plasma-fixated nitrogen (168 ppm $NO_3-N$) were used to illustrate the measurable effects of various nitrate concentrations on grass growth. Increasing the concentration of plasma-fixated nitrogen was found to increase overall height and biomass. However, nitrogen efficiency in increasing height was found to decrease with increasing nitrate concentration, with the optimal dilution for nitrogen efficiency in enhancing height being the 20:1 dilution. The optimal dilution for nitrogen efficiency in enhancing biomass seemed to be the 5:1 dilution.

Solutions of plasma-fixated nitrogen, phosphorous, and potassium were tested to investigate the interactions of the plasma-fixated nitrogen on the two other plant nutrients. Plasma-fixated nitrogen added with phosphorus and potassium were found to improve growth and dry mass compared to just plasma-fixated nitrogen and phosphorus and potassium alone. 

Finally, the effect of plasma-fixated nitrogen on grass growth were compared to two commonly used nitrate sources (potassium nitrate and sodium nitrate), and was found to have improvements in growth height and dry mass with the same dosage of nitrates (100 ppm $NO_3-N$), allowing for a more efficient nitrate fertilizer.

To summarize, we have investigated the use of plasma-fixated nitrogen as an exogenous nitrate fertilizer source for two types of commonly used turf grass. A plasma-fixated nitrogen fertilizer, which contains active nitrates ($NO_3^-$), is a promising candidate for use as a sustainable and green exogenous nitrogen source for fertilizing turf grass. 

\section*{Author Contributions}
Christina Sze: Conceptualization, Investigation, Methodology, Software Data Curation, Visualization, Writing - Original draft preparation. Benjamin Wang: Conceptualization, Investigation, Visualization, Writing - Original draft preparation. Jiale Xu: Conceptualization, Resources, Investigation. Juan Rivas-Davila - Conceptualization, Resources, Editing, Supervision, Funding acquisition. Mark Cappelli: Conceptualization, Methodology, Writing - Original draft preparation, Editing, Supervision, Funding acquisition.

\section*{Conflicts of interest}
The authors claim no conflicts of interest.

\section*{Acknowledgements}
We wish to acknowledge the Stanford TomKat Center Innovation Transfer Grant and the Stanford Wood's Institute for the Environment REIP grant for their support of this project. We would like thank Daniel James, the Stanford University Siebel Varsity Golf Training Complex Manager, for his insightful discussions and expertise on turf grass, and for providing the ryegrass seeds for our studies.

\bibliography{PAW_FINAL}

\end{document}